# INCREASING KNOWLEDGE WORKER EFFICIENCY THROUGH A "VIRTUAL MIRROR" OF THE SOCIAL NETWORK

Stori Lynn Hybbeneth,
Cisco
9155 E. Nichols Avenue,
Suite 400
Englewood CO 80112
shybbene@cisco.com

Dirk Brunnberg
Zeppelin University
Friedrichshafen,
Germany
d.brunnberg@zeppelin-university.net

Peter A. Gloor
MIT CCI
5 Cambridge Center
Cambridge MA 02138
pgloor@mit.edu

**ABSTRACT**

In this paper we introduce a case study describing the combination of manual survey-based and e-mail-based social network analysis. The goal of the project was to increase collaboration efficiency in a team of consultants of a major high tech manufacturer. By analyzing the social network of a team of 42 consultants and comparing it with their utilization as the dependent variable, their efficiency in working together was improved in various way by bridging structure holes and eliminating bottlenecks, reducing stress for overburdened individuals, connecting isolated individuals and identifying the best network structures for high utilization and increased job satisfaction.

**INTRODUCTION**

Recently social network analysis has left the academic ivory tower and has been put to productive use at the workplace. One key application is optimizing knowledge worker productivity by analyzing their organizational networks and developing interventions for improved collaboration (Bulkley & Van Alstyne 2006). In this research project we are extending previous work analyzing e-mail networks (Gloor 2006, Kidane & Gloor 2007) to analyzing one month's worth of communication archives of a team of 42 internal consultants at a globally leading high-tech manufacturer. We complement the analysis of the e-mail network with a conventional social network analysis, where the social network has been created by using a name generator.

When interviewing the project stakeholders at the beginning of the project we hypothesized that improved collaboration and the sharing of critical information between teams would increase the group's ability to scale and improve efficiencies. The goal of the project has been defined as identifying the "optimal" structure for communication, asking questions like: How does collaboration impact key metrics? How do teams work together on customer projects? What is the impact of different collaboration tools on group cohesiveness? Are there key individuals for specific types of collaboration? How can we improve collaboration to complete internal initiatives, i.e. build communities of practice? What does collaboration look like for advice and coaching, for innovation, and for trust networks?

**DATA COLLECTION**

To address these questions, we collected different types of communication archives: e-mail, bespoke enterprise social software and the interaction records in an enterprise content management system. We initiated the collection process by interviewing the managers and selected individuals in order to understand more about the mechanisms the group used to collaborate and share information with one another. As part of the study, each of the participants also completed a survey based Organizational Network Analysis (ONA) and collaboration assessment. The results of these data and survey-based analyses were compared. All of the data collected was anonymized.

**E-Mail Collection**

We were able to construct a complete social network using a subsample of mailboxes for a total of 27 of the 42 members of the group. In earlier work we had found that collecting less than twenty percent of all participants' mailboxes leads to an e-mail based social network which is over ninety percent complete (Zilli et al. 2006).

Each of the participants shared the data they had available locally and then we reviewed the common date range available across the data sets, leading to an analysis period of 3 months, from April 1st through June 30th 2012. For the purposes of this study, in order to eliminate spam, we set minimum thresholds on the number of communications for a given time period when we generated the network graphs.

**Enterprise Social Software**

An enterprise social software system was one of the newer tools made available by the organization. Social Software includes functionality such as: status updates, personal blogs, communities, ability to comment, forward and share content. Social Software is occasionally referred to as a "Facebook for Business". Understanding the level of adoption of the social software and opportunities for measuring value of the usage was of interest to the project stakeholders. Enterprise social software has unique adoption challenges within organizations. In most cases, using enterprise social software is optional, not a required aspect, of completing the day to day workflow. In fact, some individuals regard using social software as a decrease to overall productivity because of the time it takes to post and share information.

Although there are many different ways to share using an enterprise social software platform, this group had created a dedicated community for sharing and storing knowledge. The community owners and administrators were provided monthly utilization reports identifying the number of contributions (posts) and documents added to the community and the number of comments and shares for each.

**Enterprise Content Management Software**

The data from an existing content management system that the organization used to store templates for project deliverables was also included in the analysis. The content management system pre-dated the enterprise social software platform, and does not support the capability for users to collaborate around a specific piece of content. We nevertheless were able to access reports generated by this system in order to identify content contributors and those who accessed and downloaded content that was contributed.

**End User Survey**

Beyond the data driven analysis, we also conducted a multi-part Organizational Network Analysis (ONA) survey and collaboration assessment. In the ONA survey, participants were asked to identify alters for the different types of their collaborative interactions. We asked individuals to begin by thinking broadly of their network and include people with whom they actively worked with on projects, people with whom they attended meetings, people on their team, and people they connected with personally. We also gathered attribute data via the survey in order to identify the following characteristics: longevity with the organization, longevity with the group, work type (office or remote) and work location. 33 of the 42 individuals in the group completed the survey.

Once the survey respondents had identified the individuals they collaborate with, the survey copied those names into the subsequent questions in order to drill down into the significance of those relationships and the methods used for collaborating. For the former, we asked individuals to rate the frequency of collaboration with each member of their network for the following activities: collaboration for the purpose of: people finding or help getting to the right resource, collaboration, advice to solve work related problems, personal or private issues or concerns, and innovation or continuous improvement.

For the second part of the question, we asked participants to identify which platforms they used to collaborate with each identified individual. These platforms included the following: E-mail, social software, Web conferencing, Instant Messaging, Face to Face, and Video Conference Calls.

We also asked the group to complete a survey based on the book, "Collaboration" (Hansen 2009). In this survey participants were asked to rate the current versus potential value that collaboration could bring to their group in three areas: reduced costs, innovation, and revenue gains. We then asked them to also rate barriers to collaboration in the areas of: "Not Invented Here" i.e. reluctance to accept outside ideas, knowledge hoarding, search, knowledge transfer, accountability, decision making, reinforcement for collaboration and success metrics for collaboration.

These networks were subsequently compared to best practice networks, and then mirrored back to the consultants, together with steps they might take to increase their efficiency.

**RESULTS OF THE E-MAIL ANALYSIS**

We looked at the collaboration of the group of 42 at three different levels. The first level consisted of the collaboration within the core group of 42 (the "In group"). The next level looked at how the group of 42 people collaborated with the peer groups in the same area of the organization. Finally we looked at how the group of 42 collaborated across the entire enterprise (the "Out group"). In each case we set a minimum threshold for the number of communications included in the analysis in order to focus on the strongest connections (e.g. only people that exchanged at least 10 e-mails).

### In and Out Group: Cross-enterprise Collaboration

In looking at collaboration of the 42 individuals across the larger ecosystem, we adjusted the minimum communication frequency threshold to 100 and selected the top 200 nodes by betweenness centrality. From this view we found strong collaborative ties into three major areas of the organization. In each area, there was only a core number of key collaborators (i.e. the "in group"), generally not more than 10% of the group size.

### In and Out Group: Core Group of 42

For the e-mail analysis of collaboration between the core group of 42 we focused on those individuals with more than 300 communications in the time period and then further reduced the overall group size by isolating the top 100 nodes with the highest betweenness centrality. We found this core group to be a very well interconnected group overall. However, inside that group we did identify an "in" and "out" group, meaning that there was a "core within the core" – individuals with a higher betweenness centrality score than others. Although this core "in" group included the leadership team it was not exclusive of it and it did not have any correlation to the attribute data collected: location, longevity, and reporting structure.

### In and Out Group: Peer Group Relationships

Using the same view settings, we also created a view which allowed us to understand the relationships and connections the study group of 42 individuals had with their peer groups in the same area of the organization. By analyzing the betweenness centrality scores and individuals with the highest number of connections we were able to identify the peer groups with whom the group of 42 had the strongest collaborative connections, and the individuals that were key (the "in group") creating those relationships.

### Contribution Index

By calculating the Contribution Index we were able to compare the activity of individual actors as senders and receivers of messages in order to identify key roles in the team. For example, leadership roles, influencer, "expeditor" and knowledge expert roles can be recognized through the contribution index. The Actor Contribution Index Y axis measures the number of e-mails sent and received by individuals. It is defined as 1 (at the top) for individuals who only send mail, as 0 (at the middle line), when the number of communications sent and received are exactly the same, and as -1 (at the bottom) for the individuals only receiving e-mails.

The X axis shows the volume of e-mails being sent by each actor. The highlighted range in figure 2 represents the top contributors overall based on the total volume of e-mail communication. Most actors are in the lower half of the Y axis, indicating they receive more e-mail than they send, likely indicating a subject matter expert type of a role.

For instance, prior research showed that official leaders seem to be both very active and well-balanced communicators (with a contribution index close to 0) (Gloor et al. 2003).

The metric used in this study is the Average Weighted Variance of the Contribution Index (AWVCI) which is an indicator for a team's communication balance (Gloor et al. 2007). High values in AWVCI indicate that a group has some very active team members.

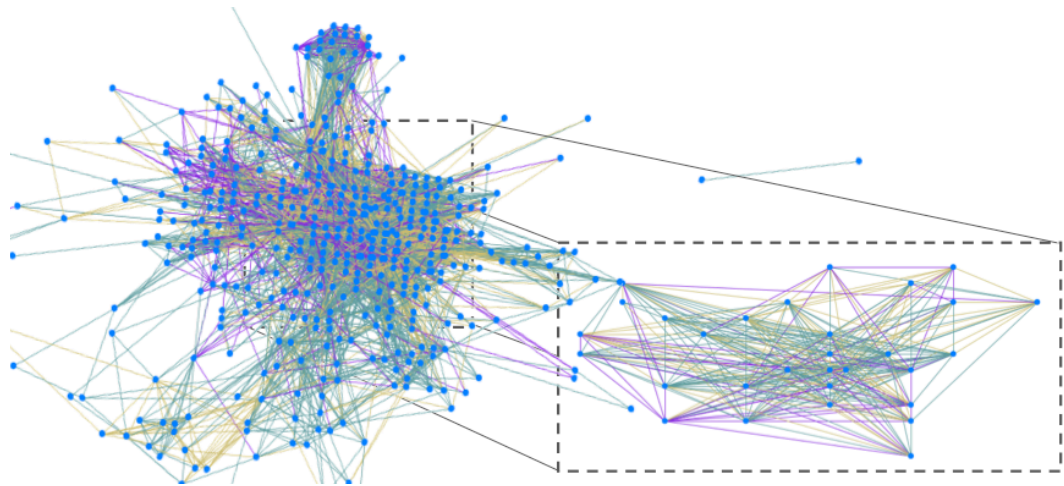

*Figure 1. In-group and out-group network (N=42)*

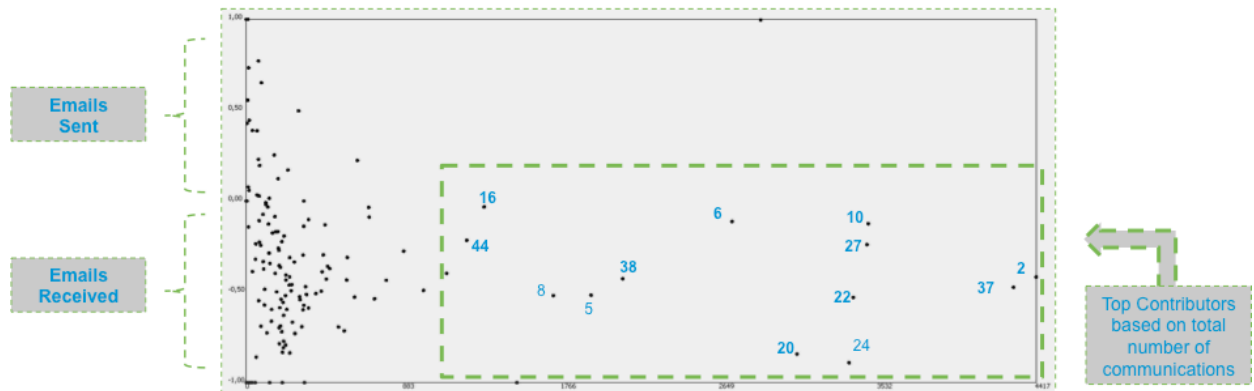

*Figure 2. Actor Contribution Index*

## Measuring Betweenness Over Time

Change in betweenness over time is an indicator of increasing and decreasing importance and changing roles of an individual. Ultimately, it also predicts individual and team creativity (Kidane & Gloor 2007). When we focused on the group of 42 individuals we found very low levels of oscillating betweenness centrality, indicating that the group was mostly execution oriented.

## Comparing the e-mail network structure with performance

As part of the initial project planning we discussed and decided upon customer facing utilization (CFU) to be used as an outcome variable. We then analyzed two months' worth of data and looked at the level of correlation between the level of collaboration and the customer facing utilization logged. This allowed us to understand which social networking structures corresponded to behavior beneficial to business.

| Core Group of 42 | 42 in full eco-system | Survey based |
|---|---|---|
| 22 | 37 | 2 |
| 16 | 22 | 37 |
| 37 | 27 | 6 |
| 27 | 16 | 27 |
| 10 | 10 | 5 |
| 6 | 33 | 7 |
| 2 | 6 | 33 |
| 33 | 2 | 16 |
| 20 | 38 | 34 |
| 26 | 44 | 38 |

*Table 2. Key individuals in all three networks (individuals common to all three groups are highlighted).*

The second part of the Organizational Network Analysis Survey analyzed the barriers to collaboration. The main barrier that was identified was "search" followed by "time".

|  | 4/1 - 4/14 | 4/15 - 4/28 | 4/29 - 5/12 | 5/13 - 5/26 | 5/27 - 6/8 | CFU - Correlations |
|---|---|---|---|---|---|---|
| **CFU** | 71,8 | 68,7 | 64,5 | 75,9 | 70,9 | |
| **Density** | 0,2381 | 0,3184 | 0,3184 | 0,2119 | 0,3129 | -0,83* |
| **Core/Periphery** | 0,2612 | 0,2027 | 0,2495 | 0,3006 | 0,2600 | 0,65 |
| **GBC** | 0,1207 | 0,0546 | 0,0546 | 0,142 | 0,0806 | 0,90* |
| **GDC** | 0,3571 | 0,345 | 0,345 | 0,4158 | 0,2941 | 0,53 |
| **AWV CI** | 0,30443984 | 0,08428862 | 0,08428862 | 0,25957334 | 0,18671094 | 0,80* |

*Table 1. Correlation between performance metric (CFU) and SNA variables (N=5)*

## Survey Based Organizational Network Analysis

The survey based organizational network analysis focused on the collaborative relationships for reaching specific purposes. Comparing the top 10 ten individuals ranked by betweenness centrality, we found that 7 of the 10 joined all three lists. Table 2 lists the top 10 individuals listed by betweenness centrality scores for each of the analyses conducted

Individuals that were identified as being key for collaboration in the area of locating and finding people and resources only slightly overlapped with the previously identified individuals with the highest levels of betweenness centrality scores.

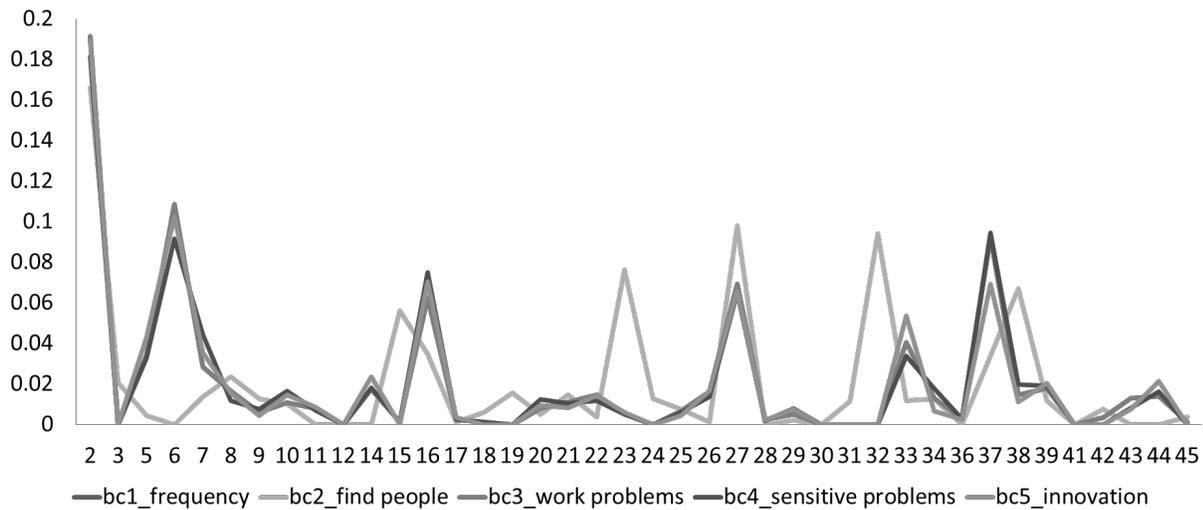

*Figure 3. Betweenness centrality of actors in the different networks obtained through survey-based ONA (e-mail, finding people, work problems, sensitive problems, innovation)*

## KEY INSIGHTS

We found that overall the study group was well interconnected. There did not seem to be individuals on the extreme periphery of the network. We also found that the network was mostly execution oriented, as opposed to exhibiting characteristics of high innovation as indicated through low oscillation in betweenness centrality.

Based on the contribution index we determined that Leadership was fulfilling their role, maintaining balanced levels of communications (sending and receiving approximately the same number of communications).

Cross-functional collaboration was restricted to a subset of the group we analyzed, it was mostly the individuals who has shared their e-mail data with us. This shows that individuals who did not see the need to participate in the study, and whose network was captured by being mentioned by others, would actually have had the biggest need to be made aware of their networking deficiencies.

As table 1 illustrates, CFU is significantly positively correlated with group betweenness centrality (GBC) and average weighted variance in contribution index (AWVCI). For any given time interval, the more centralized the networking structure, i.e. the higher GBC that means the more a few strong leaders dominate the network, the higher the group's performance. The same is true for the contribution index: the more a few actors are very active senders with the others being more passive, i.e. the higher AWVCI, the higher is group performance. The opposite is true for density: the lower the group's density, the higher is the group's performance. This means that the group performs better if actors communicate selectively.

## DISCUSSION – IT'S ALL ABOUT QUALITY!

From the survey data, we found the biggest barriers to collaboration to be search (ability to locate people and resources) and time (to collaborate with others). The most frequently used collaboration technology was e-mail, followed by Instant Messaging and Web Conferencing. We also found adoption of Enterprise Social Software and other asynchronous platforms to be minimal, indicating an area of opportunity.

Recommendations for change based on the outcome of the ONA fell into three categories: improving innovation, better sharing, and improved execution.

To improve innovation we recommended the team hold optional learning sessions where current and past projects could be reviewed. These sessions can be an open forum type of dialogue versus a presentation format. The emphasis would be placed on sharing lessons learned and having a forum for asking questions to the larger group about particular problems within a project. For ongoing access, these sessions might be conducted virtually, recorded and then posted to the community for ongoing access.

To improve quality of sharing instead of quantity of sharing we looked for opportunities to transition heavily used mailers to a discussion group. For those individuals at the core of the network, we invited them to categorize the types of requests they get from others to make information they share of common interest more widely available, such as posting to the community site or announcing it in a group forum.

We also suggested a mentorship program for new people. Pairing core and periphery individuals would help them navigate the network and build ties.

Although the core group included members of the leadership team, it was not exclusive to leadership, we also did not find any other correlating attributes.

Finally, we recommended increasing reinforcement in the form of recognition and rewards for the creation and sharing of project files, templates, lessons learned, etc.

To improve execution we recommended the group take more advantage of the video streaming capabilities already available. Although they were able to share video over their web conference platform, many did not. The corporate culture supported employees developing trust environments through visual connection with the belief being as trust develops; the team is able to execute more quickly. Of course, this may mean that the social norms and expectations of how people look while working will change.

In sum, we found that it's really quality of communication that matters. Just spamming team members with information consumes valuable "information processing cycles." The people in highest demand need to transfer some of their knowledge into more permanent repositories than e-mail. While one can never share enough knowledge, it really matters how it is done, we envision a future where there is much more information pull than push – but this requires a paradigm shift in sharing behavior.